\documentclass[12pt]{article}

\usepackage{amsmath,amsfonts,amssymb}

\newcounter{tabl}
\newcommand{\be}{\begin{equation}}
\newcommand{\ee}{\end{equation}}
\newcommand{\beq}{\begin{eqnarray}}
\newcommand{\eeq}{\end{eqnarray}}
\newcommand{\bea}[2]{\be\label{#2}\begin{array}{#1}}
\newcommand{\eea}{\end{array}\ee}

\def\Cb{{\rm \bf C}}
\def\Zb{{\rm \bf Z}}
\def\Tr{\,{\rm Tr}\, }
\def\det{\,{\rm det}\, }

\def\tr{\,{\rm tr}\, }

\def\({\left(}
\def\){\right)}
\def\[{\left[}
\def\]{\right]}
\def\p{\partial}

\def\11{1\!\! 1}

\def\hf{{1\over 2}}

\def\eps{\varepsilon}

   \def\CD {{\cal D}}

   \def\CI {{\cal I}}

   \def\CO {{\cal O}}

   \def\CV {{\cal V}}

\newcommand{\SA}{{\cal A}}
\newcommand{\SSA}{{\bf A}}

\newcommand{\gx}{{\rm g}}
\newcommand{\gl}{g}
\newcommand{\gb}{{\rm g}}
\newcommand{\Db}{{\bf D}}

\newcommand{\gh}{g^{\hphantom{1}}}
\newcommand{\xh}{x^{\hphantom{1}}}
\newcommand{\rhoh}{\rho^{\hphantom{1}}}

\newcommand{\cI}{\CI^{(\lambda_i)}}
\newcommand{\cIr}{\CI^{(\lambda_i)}_{\rm rel}}

\newcommand{\ClG}[6]{{{C\lefteqn{\vphantom{\overline{\hat A}_-}}^{\smash{#1}}_{\smash{#4}}}^{\smash{#2}}_{\smash{#5}}}^{{#3}}_{{#6}}}

\begin{document}
%
%

\title{
Simplicity and closure constraints \\ in spin foam models of gravity}

\author{Sergei Alexandrov\thanks{email: Sergey.Alexandrov@lpta.univ-montp2.fr}
}

\date{}

\maketitle

\vspace{-1cm}

\begin{center}
\emph{Laboratoire de Physique Th\'eorique \& Astroparticules} \\
\emph{Universit\'e Montpellier II, 34095 Montpellier Cedex 05, France}


\end{center}

\vspace{0.1cm}

\begin{abstract}

We revise imposition of various constraints in spin foam models of 4-dimensional general relativity.
We argue that the usual simplicity constraint must be supplemented by a constraint
on holonomies and together they must be inserted explicitly into the discretized path integral.
At the same time, the closure constraint must be relaxed so that
the new constraint expresses {\it covariance}
of intertwiners assigned to tetrahedra by spin foam quantization.
As a result, the spin foam boundary states are shown to be realized in terms of projected spin networks
of the covariant loop approach to quantum gravity.

\end{abstract}

\thispagestyle{empty}

\newpage

\pagenumbering{arabic}

\tableofcontents

\vspace{0.4cm}

\section{Introduction}

The spin foam approach is a way to quantize gravity in a spacetime covariant setting \cite{Oriti,perez}.
Whereas in 3 dimensions, where gravity is a particular case of the so called BF theory,
the rules of spin foam quantization and the geometric interpretation of results
are more or less clear \cite{PonzanoRegge,Noui:2004iy},
the physically relevant case of 4 dimensions is still an arena for debates.
One of the main non-settled issues is how to implement the constraints which appear
in the formulations of 4-dimensional general relativity used in the spin foam quantization.
This is the problem which we are going to address in this paper.

\subsection{Simplicity constraints}

The classical formulation of general relativity, which is the most suitable to build spin foam models,
was suggested by Plebanski \cite{Plebanski:1977zz}. It represents gravity as a simple topological
BF theory supplemented by the so called {\it simplicity constraints}.
These constraints are conditions on the $B$ field of BF theory
that ensure that it is constructed from tetrad one-forms
\be
B^{IJ}=*(e^I\wedge e^J),
\label{Bee}
\ee
where $*$ is the Hodge operator acting on tangent space indices.
The constraints can be written in the following form
\be
\eps_{IJKL} B^{IJ}_{\mu\nu} B^{KL}_{\rho\sigma}=\sigma\,
{\cal V} \, \eps_{\mu\nu\rho\sigma} ,
\label{simplicityconditions1}
\ee
where $\CV= \frac{1}{4!}\,\eps_{IJKL}\eps^{\mu\nu\rho\sigma} B_{\mu\nu}^{IJ} B_{\rho\sigma}^{KL}$ is the 4-volume
and $\sigma=\pm 1$ in the Riemannian/Lorentzian case.

The standard strategy of the spin foam quantization is first to quantize the BF theory,
and then to impose the simplicity constraints already at the level of the discretized path integral
\cite{De Pietri:1998mb,Freidel:1998pt}. In particular, this means that the constraints
are imposed not directly on the $B$ field, since it is integrated out, but on representations and
intertwiners of the gauge group which are assigned to elements of a simplicial decomposition of spacetime.
These group theoretic data are thought as quantum degrees of freedom corresponding to the $B$ field.
As a result, it is a hard task to find a consistent and correct way implementing the constraints.

For a long time the most popular model in 4 dimensions was the one proposed by Barrett and Crane (BC)
\cite{BCE,BC}. However, it has become clear that it is insufficient to describe the genuine general relativity.
Therefore, recently various modifications of the Barrett--Crane quantization procedure
have been suggested to improve the BC model
\cite{Livine:2007vk,Engle:2007uq,Alexandrov:2007pq,Engle:2007qf,Freidel:2007py,Livine:2007ya,Oriti:2007vf,Engle:2007mu,Pereira:2007nh,Engle:2007wy}.
All these proposals differ from the original model and between each other in the way
they treat the simplicity constraints.

Note that some of the new models suggest a way to overcome the above mentioned problem of disappearance
of the $B$ field in the quantum partition function.
In particular, the approach based on coherent states \cite{Livine:2007vk,Freidel:2007py,Livine:2007ya}
allows to give a faithful representation for the $B$ field at the quantum level in terms of
additional geometric data appearing as labels of these states.
Therefore the imposition of the simplicity constraints in this approach
becomes much more transparent and reliable comparing to previous studies.
Another interesting development was done in \cite{Oriti:2007vf} where a new class of
group field theories was proposed. These models keep the $B$ field in
the partition function explicitly which might be crucial for the correct implementation of
the constraints.

There is however one additional complication which is usually neglected in most of
the old and new spin foam models.
As it is clear from canonical analysis of Plebanski formulation \cite{Buffenoir:2004vx,ABR},
the simplicity constraints are supplemented by secondary constraints involving the gauge connection.
Altogether they form a system of second class constraints. The question now is: should these
secondary constraints be inserted explicitly into the path integral?

One usually assumes that one can start from the Lagrangian path integral with a trivial measure,
which does not involve contributions from any constraints. But it is generally believed that
the phase space path integral is more fundamental. The latter certainly contains delta-functions
of all constraints and a non-trivial measure. Therefore, whether the simple Lagrangian path integral
can be used depends on our possibility to derive it from the phase space path integral. Whereas
there is a general method for this purpose \cite{Henneaux:1994jf}, it turns out that it has some
restrictions and should be used with care. In particular, we will argue
that the secondary constraints of Plebanski formulation cannot be removed and must be taken into
account in the spin foam quantization.

The latter requires however to realize the secondary constraints at the discretized level
in terms of holonomies and bi-vectors. Unfortunately, we were not able to find such a realization
for the constraints relevant for the gravitational sector of Plebanski formulation.
Nevertheless, it is possible to show that it is these constraints that are responsible
for the restrictions on representations and intertwiners of a simplicial decomposition.
We demonstrate also how this leads to constructions appearing in
the framework of covariant loop quantum gravity (CLQG) \cite{SA,AV,SAcon,AlLiv,Livine:2006ix}.

\subsection{Closure constraint}

Besides the simplicity constraints, the spin foam quantization involves another type of constraints,
{\it the closure constraint}. The latter appears in all spin foam models and represents the Gauss constraint
of canonical formulation realized in the discrete setting. Thus, it ensures the fundamental gauge invariance
of quantized theory.

The closure constraint can be formulated as follows.
Let us consider a simplicial decomposition of spacetime which the spin foam quantization is based on.
In 4 dimensions such decomposition is formed by
4-simplices, tetrahedra, triangles, segments and points. The triangles are colored with irreducible representations
of the relevant gauge group $G$, which will be taken either SO(4) or SL(2,$\Cb$),
and the tetrahedra are equipped with intertwiners between the representations
assigned to its sides.
We label the triangles and the tetrahedra by $f$ and $t$, respectively.
If we denote the representation assigned to the $f$th triangle by $\lambda_f$,
then the intertwiner $N_t$ is a vector in the tensor product of four representation spaces,
$N_t\in {\bigotimes_{f\subset t}}\, H_G^{(\lambda_f)}$.
The closure constraint requires that the intertwiners are gauge invariant, {\it i.e.}
\be
\vphantom{A\over B}
\smash{\sum_{f\subset\, t}}\, T_f\, N_t=0,
\label{clcon}
\ee
where $T_f$ are generators of the gauge algebra acting on $H_G^{(\lambda_f)}$.
In this form the constraint was widely used in various constructions including the original
BC model and the recent proposals \cite{Engle:2007uq,Engle:2007qf,Engle:2007mu,Pereira:2007nh,Engle:2007wy}.

However, in the work \cite{Alexandrov:2007pq} a relaxed version of this constraint has been proposed.
It can be written as
\be
\vphantom{A\over B_B}
\smash{\sum\limits_{{f\subset\, t}}}\, T_f\, N_t(x_t)=\hat T\cdot N_t\(x_t\),
\label{clconrel}
\ee
where $x_t\in G/H$ ($H$ is the maximal compact subgroup of $G$) is interpreted as normal to the tetrahedron
and $\hat T$ acts on functions of $x_t$ as a generator of $G$
\be
\hat T_{IJ}\cdot f(x)=\eta_{IK} x^K \p_J f-\eta_{JK} x^K \p_I f.
\label{generator}
\ee
The intertwiners satisfying \eqref{clconrel} are not invariant anymore, but rather covariant.
The relaxed constraint was motivated by comparison with results of canonical quantization
obtained in the framework of CLQG.
The use of the constraint \eqref{clconrel} together with a modified identification
of bi-vectors with generators of the gauge algebra ensured the coincidence of boundary states
of the resulting spin foam model with kinematical states of CLQG \cite{Alexandrov:2007pq}.

In this paper we justify the new constraint \eqref{clconrel} from the pure
spin foam point of view. The key point for this is the consistent implementation of second class constraints
and, in particular, the secondary constraints mentioned above.
In the discrete setting these constraints involve the normals to tetrahedra.
It turns out that this fact requires the normals $x_t$ to be considered as
additional arguments of the boundary states and leads precisely to the intertwiners satisfying \eqref{clconrel}.
The usual invariant intertwiners \eqref{clcon} can be obtained by integrating over the normals $x_t$,
what however would make impossible implementation of the simplicity and secondary constraints.

\

The organization of the paper is as follows.
In the next section we discuss the general method to pass from
the phase space path integral to the Lagrangian one in the presence of second class constraints
and argue that it does not work for generic correlation functions.
In section \ref{sec_simconstr_new} we revise the derivation of the simplex
boundary state and express it in terms of projected spin networks of CLQG.
Then in section \ref{sec_conseq} we discuss
implications of our approach on the closure constraint and demonstrate that only its relaxed version
is consistent with the second class constraints.
Finally we consider an example which allows to make various aspects of the construction explicit.

\section{Path integral in presence of second class constraints}
\label{sec_constraint}

Let us consider a dynamical system with the phase space parameterized by $(q^a,p_a)$ and
subject to second class constraints. For the sake of simplicity, we suppose that there are only
two constraints one of which, say $\phi$, is primary  and the second, $\psi$, is secondary.
This means that
\be
\psi=\{ H, \phi\}, \qquad \{ \phi,\psi\}\ne 0,
\ee
where $H$ is the Hamiltonian.

Correlation functions for this system are defined by the phase space path integral
\be
\langle\CO\rangle = Z^{-1}\int \CD q\,\CD p \, |\det\{\phi,\psi\}|\,\delta(\phi)\delta(\psi)\, e^{i\int dt\(p_a \dot q^a-H\)}\CO(q,p),
\label{phspint}
\ee
where $\CO$ is an observable and $Z$ is the partition function.
The problem we would like to address is whether these correlation functions are equal to their analogues
defined by the configuration space path integral. Since we want to include into consideration also systems
with first order Lagrangians (like Plebanski or Palatini formulations of gravity), by latter we mean
\be
\langle\CO\rangle_{\rm c.s.} =
Z_{\rm c.s.}^{-1}\int \CD q\,\CD p \, \CD\lambda\, \rho(q,p)\, e^{i\int dt\(p_a \dot q^a-H-\lambda\phi\)}\CO(q,p),
\label{confspint}
\ee
where $\lambda$ is the Lagrange multiplier for the primary constraint, $\rho(q,p)$ is a regular local measure,
and the expression in the exponential is the Lagrangian in its first order formulation.
Comparison of \eqref{phspint} and \eqref{confspint} shows that the difference between the two expressions
arises due to the secondary constraint $\psi$. To bring the correlation functions together,
one should somehow remove its contribution.

In \cite{Henneaux:1994jf} an elegant method has been suggested to achieve this goal for the partition function.
It makes use of the canonical transformation generated by $\mu\phi$ where $\mu$ is the Lagrange multiplier
which can be introduced for the secondary constraint.
Implemented in the action
$\int dt\(p_a \dot q^a-H-\lambda\phi-\mu\psi\)$,
it cancels the term linear in $\mu$ and therefore the integral over this Lagrange multiplier produces a regular
local measure instead of the $\delta$-function of the constraint.

However, this idea does not work so well for the correlation functions because the phase space function
$\CO(q,p)$ gets transformed together with the integral. As a result, one obtains a correlation function
of the following form
\be
\langle\CO\rangle = Z^{-1}\int \CD q\,\CD p \,\CD \lambda \, \CD\mu \, |\det\{\phi,\psi\}|\,
e^{i\int dt\(p_a \dot q^a-H-\lambda\phi-\hf\,\mu^2\{\phi,\psi\}+O(\mu^3)\)}
\CO'(q,p,\mu),
\label{phspint_tr}
\ee
where $\CO'$ is the result of the canonical transformation of the initial function $\CO(q,p)$.
Even if one succeeds to perform the integral over $\mu$, the result will differ from \eqref{confspint}.
Roughly speaking, one will obtain the correlation function of a different observable.
Thus, the two definitions, \eqref{phspint} and \eqref{confspint}, do not coincide in general.

This situation can be illustrated on very simple examples.
The simplest example is provided by the trivial system with the Lagrangian $L=\hf\, q^2$.
It is clear that it corresponds to the situation considered above with $\phi=p$, $\psi=q$.
For this system $\langle q^2\rangle_{\rm c.s.}>\langle q^2\rangle=0$. The reason for the inequality can easily be traced back
to the non-commutativity of the observable $\CO=q^2$ with the primary constraint $\phi$
used in the canonical transformation so that in this case $\CO'=(q-\mu)^2$.

A similar example is given by the Lagrangian $L=p\dot q -\hf\, p^2-\lambda q$, which leads to
the constraints $\phi=q$ and $\psi=-p$. Now the two definitions do not coincide
for the correlation functions of observables dependent on the momentum $p$.

It is clear that in both examples the canonical quantization based on the introduction of a Dirac bracket
would lead to results consistent with the phase space path integral.
Thus, we conclude that if one wishes to consider correlation functions of observables dependent on all
phase space variables, one has to work with the path integral where all constraints, primary and secondary,
appear in $\delta$-functions.\footnote{We remark also that the modern approach to the Lagrangian path integral
originated from the works of Batalin and Vilkovisky \cite{Batalin:1981jr,Batalin:1984jr}
leads to similar conclusions \cite{Batalin:1997dy}.}

\section{Simplicity constraints revisited}
\label{sec_simconstr_new}

\subsection{Constraints on connection}
\label{subsec_connect}

The main lesson of the previous section is that one should not ignore the secondary second class constraints.
Four dimensional gravity in the first order formulation is an example of the system possessing such type
of constraints.
Therefore, quantizing gravity by the spin foam approach, one must include
them into the measure together with the usual simplicity constraints.
Let us briefly recall what the canonical analysis tells us about the secondary constraints of Plebanski formulation.

In fact, in \cite{ABR} it has been shown that the canonical formulation of Plebanski action \cite{Buffenoir:2004vx}
is essentially equivalent to the Lorentz covariant canonical formulation of the Hilbert--Palatini action \cite{SA}.
Therefore, we can use the results from the latter where the constraints have been extensively studied.

The secondary constraints, which appear by commuting the Hamiltonian with the simplicity constraints,
depend linearly on the spin connection. This connection however is not suitable for quantization
and it is more convenient to consider another connection, which we call $\SA_i^{IJ}$,
differing from the original one by a shift by the Gauss constraint \cite{AV}.
Its holonomies have simple commutation relations with the smeared triad what has important consequences
for quantum theory. Due to this, it is advantageous to formulate the secondary second class constraints also
in terms of this new variable. They have the following form
\be
I_{(Q)KL}^{IJ}\SA_i^{KL}=\Gamma_i^{IJ}(B),
\label{contA}
\ee
where $\Gamma_i^{IJ}$ is the Levi--Civita connection determined by the space components of the $B$ field
and we introduced two projectors
\be
I_{(Q)}^{IJ,KL}(x)=
\eta^{I[K}\eta^{L]J}-2\sigma\, x^{[J}\eta^{I][K} x^{L]},
\qquad
I_{(P)}^{IJ,KL}(x)=2\sigma\,x^{[J}\eta^{I][K} x^{L]}.
\label{projxxx}
\ee
They depend on a 4-dimensional unit vector $x^I$ ($x^I x_I=\sigma$), which defines the normal to three-dimensional
hypersurfaces foliating spacetime and it can be extracted from the $B$ field by solving the simplicity
constraints. This vector defines a subgroup $H_x={\rm SU}_x(2)$ of the gauge group $G$
which leaves it invariant. Then the geometric meaning of $I_{(Q)}^{IJ,KL}$ and $I_{(P)}^{IJ,KL}$ is
that they project on the algebra of $H_x$ and its orthogonal completion, respectively.
These projectors play an important role and appear also in the symplectic structure
determining the commutator of the shifted connection with the $B$ field
\be
[\, \eps^{jkl}B_{kl}^{IJ}(y),\SA_i^{KL}(x)]
=-i\hbar\,I_{(P)}^{IJ,KL}\delta_i^j\delta(x,y).
\label{commPleb}
\ee

Given the geometric interpretation of the projectors, it is clear that the meaning of the constraints
\eqref{contA} is to restrict the ``rotational" part of the shifted connection as a function of
the $B$ field.\footnote{A reader
familiar with the self-dual Ashtekar formulation of general relativity \cite{Ashtekar:1986yd,Ashtekar:1991hf}
may find similarities between the constraints \eqref{contA} and the so called reality conditions imposed on complex
Ashtekar connection \cite{Ashtekar:1991hf,Immirzi:1992ar}. And indeed the reality conditions can be shown to be
equivalent to our second class constraints \cite{Alexandrov:2005ng}.}
The ``rotational" part is defined by the vector $x^I$ as explained above.
Therefore, all independent physical degrees of freedom of the connection are contained only in its ``boost" part.

It is worth to notice that one can quantize the theory using another object, $\SSA_i^{IJ}$, instead of the shifted connection
$\SA_i^{IJ}$. This object has all properties of a gauge
connection except that it does not transform appropriately under time diffeomorphisms.
It has an advantage that it is commutative and in the time gauge $x^I=\delta^{I,0}$ the loop quantization based on
holonomies defined by $\SSA_i^{IJ}$ reproduces the results of LQG \cite{SAcon,AlLiv}. This can be traced back to the secondary constraints
which in terms of $\SSA_i^{IJ}$ become
\be
I_{(P)KL}^{IJ} \SSA_i^{KL}=2\,x^{[J}\p_i x^{I]},
\label{su2con}
\ee
They imply that in the time gauge $\SSA_i^{IJ}$ becomes an SU(2) connection coinciding with the one used
in the Ashtekar--Barbero approach \cite{Ashtekar:1986yd,Barbero:1994an}. However, the failure of this quantity to
have correct spacetime transformations points in favor of quantization based on the true spacetime connection $\SA_i^{IJ}$.
It is the latter quantization that gives rise to the CLQG approach.

\subsection{Discretization}
\label{subsec_discr}

Before quantizing gravity in the spin foam approach, one needs to realize its degrees of freedom at the discrete level.
Given a simplicial decomposition of spacetime, one considers the following data:

i) elements of the gauge group $\gl_{\sigma t}$ assigned to every couple of simplex $\sigma$ and tetrahedron $t$, which
represents holonomy of the gauge connection from the center of the former to the center of the latter;

ii) bi-vectors $B_f^{IJ}$, which can be thought as elements of the gauge algebra $\hat B_f=B_f^{IJ} T_{IJ}$,
assigned to every triangle $f$.

\noindent
These data are enough to encode all degrees of freedom and to represent the gravity action at the discretized
level (see, for example, \cite{Engle:2007qf}). However, one has to also rewrite
the second class constraints in terms of the discrete variables $B_f^{IJ}$ and $\gl_{\sigma t}$.

The discretization of the quadratic simplicity constraints \eqref{simplicityconditions1} has been extensively
studied in the literature. There is however a nice way to effectively linearize them. It makes use
of the projectors \eqref{projxxx}. It turns out that the simplicity constraints can be rewritten as \cite{Alexandrov:2007pq}
\be
I_{(Q)KL}^{IJ}(x)B^{KL}=0.
\label{newsim}
\ee
Although the constraint \eqref{newsim} seems to be linear, one should remember that $x$ is a part
of the $B$ field. What is really going on here is a decoupling of degrees of freedom: once $x$ has been extracted,
the rest of the $B$ field is constrained to satisfy \eqref{newsim} and the only remaining freedom
corresponds to the triad of canonical theory.

At the discrete level the $B$ field gives rise to bi-vectors normal to triangles.
Taking into account that $I_{(Q)}(x)$ annihilates bi-vectors coaligned with $x^I$,
it is easy to understand the geometric meaning of the condition \eqref{newsim}. It requires that
all triangles, for which the constraint is written with the same $x^I$, must lie in the hypersurface
normal to this 4-dimensional vector. Therefore, $x^I$ is naturally identified with the normals to tetrahedra
of the simplicial decomposition.
Regarding this, we introduce an additional set of geometric data:

iii) unit vectors $x_t^I$ playing the role of normals to tetrahedra;

iv) elements of the gauge group $\gx_{ft}$ assigned to every couple of tetrahedron $t$ and triangle $f$
and constrained to satisfy $\gx_{ft}\cdot x_t=\gx_{ft'}\cdot x_{t'}$ for two tetrahedra sharing $f$.

\noindent
The introduction of $\gx_{ft}$ is related to the subtlety arising after discretization that
the normals $x_t$ and the bi-vectors $B_f$ all live in different frames. Therefore,
we need to parallel transport them to be able to compare. The elements $\gx_{ft}$ just serve
to this purpose. They are not really important and will disappear from the action and the path integral.

Given these additional data, it is now easy to formulate the simplicity constraints at the discrete level.
Let us define the operator
\be
\hat I_{(Q)t}^{IJ} = I_{(Q)}^{IJ,KL}(x_t) T_{KL}.
\ee
Then the simplicity constraints \eqref{newsim} become
\be
\phi_{f}^{IJ}=\Tr \( \hat I_{(Q)t}^{IJ}\, \gx_{ft}^{-1} \, \hat B_f \, \gx_{ft}^{\hphantom{1}} \).
\label{opsim}
\ee
Due to the condition on $g_{ft}$, the r.h.s. does not depend on the chosen tetrahedron
and therefore we omitted the label $t$.

The secondary constraints \eqref{contA} are much more difficult to discretize.
The reason is that, whereas the simplicity constraints are formulated at the algebra level,
the constraints on connection should be realized as a condition on group elements.
In this paper we leave this problem unsolved. Nevertheless, it is possible to draw some important conclusions
just assuming that there is a measure on the space of holonomies incorporating the secondary constraints.
For this we will need only one property of this measure.

As we saw in the previous subsection, all versions of the secondary constraints, similarly to the simplicity
constraints \eqref{newsim}, depend on $x^I$.
This implies that the discrete measure should contain a dependence of the normals to tetrahedra and, what is important,
it should appropriately transform under the gauge group.
Thus, we assume that there exists a measure $\CD^{(x_t)} [\gl_{\sigma t}]$ on the space of holonomies
dependent of the normal $x_t$. It may also have a dependence of $B_f$ which we do not display explicitly.
The measure is required to transform covariantly with respect to the group
transformations\footnote{Here the measure is defined for holonomies between a simplex and one of its tetrahedra.
This is the reason why only the transformation property under the right group action is imposed. To define the left group action,
one would need to introduce a 4-dimensional vector associated with a 4-simplex. But there are no such natural vectors.
In any case, the holonomies always appear in combinations $\gl_{\sigma t}^{-1} \gl_{\sigma t'}$ and therefore the group transformations acting
from the left are not important.}
\be
\CD^{(x)} \[\gl\, \gb\]=\CD^{(x^\gb)} [\gl], \qquad x^\gb=\gb\cdot x.
\label{trmeas}
\ee
An important consequence of \eqref{trmeas} is that the measure defined by
the normal fixed to $x_0=(1,0,0,0)$ is invariant with respect to the diagonal SU(2) subgroup
\be
\CD^{(x_0)} \[\gl \,\gb^h\]=\CD^{(x_0)} [\gl], \qquad \gb^h=(h,h).
\label{invmeas}
\ee

Finally, the unconstrained action at the discrete level reads
\be
S_{BF}=\sum_f \Tr \( \gx_{ft_1}^{-1} \, \hat B_f \, \gx_{ft_1}^{\hphantom{1}} \,
\gl_{\sigma_{12} t_1}^{-1}\gl_{\sigma_{12} t_2}^{\hphantom{1}}
\cdots \gl_{\sigma_{n 1} t_n}^{-1}\gl_{\sigma_{n1} t_1}^{\hphantom{1}} \),
\label{dact}
\ee
where the trace includes the product over all tetrahedra containing a given triangle and this product
does not depend on the ``reference" tetrahedron $t_1$.
Comparing with \eqref{opsim}, we see that choosing such a ``reference" tetrahedron $t_{f}$ for every triangle, it is natural to redefine
$\gx_{ft_{f}}^{-1} \, \hat B_f \, \gx_{ft_{f}}^{\hphantom{1}}\to \hat B_f$ so that the dependence of auxiliary
group elements $g_{ft}$ disappears.

\subsection{Simplex boundary state}
\label{subsec_sbs}

Now we would like to reconsider the derivation of the simplex boundary state taking into account
the previous results. Thus, we include the contribution from both primary and secondary second class constraints
into the measure from the very beginning. This means in particular that we give up the usual strategy used
in the spin foam quantization: first quantize and then impose constraints.

The natural object to start with is the discretized path integral for a single 4-simplex
with fixed $B_f$ on the boundary. Inclusion of all constraints amounts to inserting the $\delta$-function
of $\phi_{f}^{IJ}$ \eqref{opsim} and taking the measure $\CD^{(x_t)} [\gl_t]$ for holonomies. This gives
\beq
A[B_f]&=&\int \prod_t \CD^{(x_t)} [\gl_t]
\prod_f\[\delta\(\phi_{f} \)
e^{i \Tr\(B_f \gl_{u(f)}^{-1}\gl_{d(f)}^{\hphantom{1}}\)}\]
\nonumber \\
&=&\int \prod_f \[ d g_f\, e^{i\Tr(B_f g_f)}\delta\(\phi_{f}\)\]
\int\prod_t \CD^{(x_t)} [\gl_t] \prod_f\delta\( \gl_{u(f)}^{\hphantom{1}}g_f\gl_{d(f)}^{-1}\),
\label{amplitgen}
\eeq
where $u(f)$ and $d(f)$ denote two tetrahedra which share the $f$th triangle. Since the triangle is oriented,
one of them is considered as ``up" and the other as ``down".
From \eqref{amplitgen} one gets the simplex boundary state in the ``connection" representation
\be
A[g_f;x_t]=\int\prod_t \CD^{(x_t)} [\gl_t]
\prod_f\delta\( \gl_{u(f)}^{\hphantom{1}}g_f\gl_{d(f)}^{-1}\).
\label{simbs_init}
\ee
Notice that the amplitude depends on the group elements $g_f$, playing the role of external holonomies,
as well as on the normals $x_t$.\footnote{If there is a dependence of $B_f$
in the measure $\CD^{(x_t)} [\gl_t]$, one can formally replace all $B_f$
by the functional derivative $\frac{1}{i}\,\frac{\delta}{\delta g_f}$.}
The latter dependence could be removed by an explicit integral over $x_t$. However,
it would be inconsistent with gluing of different simplices and the proposed measure.
We refer to section \ref{sec_conseq} for more detailed discussion of this important issue
which has direct consequences for the closure constraint and the boundary state space.

The normals $x_t$ can be considered as elements of the homogeneous factor space $G/H$.
It will be convenient to denote by $\gx_x$ a representative of $x\in X$ in $G$ so that $\gx_x \cdot x_0=x$.
Changing the variable $\gl_t\to \gl_t\gx_{x_t}^{-1}$, one can use the covariance property \eqref{trmeas}
to extract the dependence of the normals from the measure
\beq
A[g_f;x_t]&=&
\int\prod_t \CD^{(x_0)} [\gl_t]
\prod_f\delta\( \gl_{u(f)}^{\hphantom{1}}G_f\gl_{d(f)}^{-1}\)
\nonumber
\\
&=&  \sum_{\lambda_f}\int\prod_t \CD^{(x_0)} [\gl_t]
\prod_f d_{\lambda_f}\tr_{\lambda_f}\( \gl_{u(f)}^{\hphantom{1}}G_f\gl_{d(f)}^{-1}\),
\label{vertgen}
\eeq
where we denoted $G_f=\gx_{x_{u(t)}}^{-1}\, g_f\, \gx_{x_{d(t)}}^{\hphantom{1}}$,
$d_{\lambda}$ is the dimension of the representation $\lambda$
and we used the Plancherel decomposition of the $\delta$-function on the group into the sum over
irreducible representations $\lambda_f$.\footnote{In the Lorentzian case $\sum_{\lambda}d_{\lambda}$
must be replaced by an integral over the spectrum of irreducible representations with the Plancherel measure.}
To proceed further, one can insert an additional integral over the diagonal subgroup SU(2) with group elements inserted
between $\gl_t$ and $G_f$. This insertion does not change the amplitude
because the group elements are absorbed into $g_t$ and disappear due to the invariance property
of the measure \eqref{invmeas}.
This gives
\be
A[g_f;x_t]=\sum_{\lambda_f}\int\prod_t \CD^{(x_0)} [\gl_t]\int \prod_t dh_t
\prod_f d_{\lambda_f}\tr_{\lambda_f}
\( \gl_{u(f)}^{\hphantom{1}} \gb^{h_{u(f)}} G_f (\gb^{h_{d(f)}})^{-1} \gl_{d(f)}^{-1}\).
\label{simbs_inter}
\ee
The following computations will be done assuming that we are working in the Riemannian case.
But the final conclusions will be true for the Lorentzian model as well.

For the Riemannian signature the relevant gauge group $G={\rm SO}(4)$
is the product of two SU(2) groups (factorized by $\Zb_2$), and
its irreducible representations are labeled by two SU(2) spins $\lambda=(j^+,j^-)$.
We will represent an element $g\in{\rm SO}(4)$ as $(g^+,g^-)$, $g^\pm\in{\rm SU}(2)$.
Thus, in \eqref{simbs_inter} one encounters group integrals of eight matrix coefficients
that can be represented as follows
\be
\int dh_t \prod_{f_i\in \,t} D^{(j^+_{f_i})}_{m_i n_i^{\vphantom{-}}}(h_t)D^{(j^-_{f_i})}_{m'_i n'_i}(h_t)
=\prod_{f_i\in\, t}\(\sum_{j_{tf_i}} \ClG{j^+_{f_i}}{j^-_{f_i}}{j_{tf_i}^{\vphantom{+}}}{m_i^{\vphantom{+}}}{m'_i}{\ell_{f_i}}
\overline{\ClG{j^+_{f_i}}{j^-_{f_i}}{j_{tf_i}^{\vphantom{+}}}{n_i^{\vphantom{+}}}{n'_i}{l_{f_i}}} \)
\sum_{k_t}\iota^{\{ j_{tf_i}\} }_{\{ \ell_{f_i} \} }(k_t)\overline{\iota^{\{ j_{tf_i}\} }_{\{ l_{f_i} \} }(k_t)},
\ee
where $C\lefteqn{\vphantom{\overline{\hat A}_-}}^{{j_1}\ {j_2}\ {j_3}}_{{m_1}{m_2}{m_3}}$ are SU(2) Clebsch-Gordan coefficients
and $\iota^{\{ j_i\} }_{\{ m_i \} }(k)$ are matrix elements of an invariant SU(2) intertwiner between 4 representations
with spins $j_i$, $i=1,\dots,4$, and with intermediate spin $k$.
The sum over repeated representation indices is implied.
Thus, the integral over $h_t$ yields the amplitude in the following factorized form
\be
A[g_f;x_t]=\sum_{\lambda_f,j_{tf},k_t}\(\prod_f d_{\lambda_f}\)A(\lambda_f,j_{tf},k_t)\Psi^{(\lambda_f,j_{tf},k_t)}[g_f;x_t] ,
\label{reprampl}
\ee
where
\be
\Psi^{(\lambda_f,j_{tf},k_t)}[g_f;x_t]=\prod_f\(
\overline{\ClG{j^+_f}{j^-_f}{j_{u(f)f}^{\vphantom{+}}}{m^{\vphantom{+}}}{m'}{\ell_{u(f)f}}}
D^{(j^+_f)}_{mn^{\vphantom{+}}}(G^+_f)D^{(j^-_f)}_{m'n'}(G^-_f) \ClG{j^+_f }{j^-_f }{j_{d(f)f}}{n^{\vphantom{+}}}{n'}{\ell_{d(f)f}} \)
\prod_t \iota^{\{ j_{tf_i}\} }_{\{ \ell_{tf_i} \} }(k_t),
\label{projspin}
\ee
\be
A(\lambda_f,j_{tf},k_t)=\int \prod_t \CD^{(x_0)} [\gl_t]\,
\Psi^{(\lambda_f,j_{tf},k_t)}\[\gl_{d(f)}^{-1}\gl_{u(f)}^{\hphantom{1}};x_0\].
\label{amplit_simplex}
\ee

It is easy to realize that the function
$\Psi^{(\lambda_f,j_{tf},k_t)}[g_f;x_t]$ is the so called projected spin network. Such states have been introduced in the
context of CLQG \cite{psn,AlLiv} and form the (enlarged) kinematical Hilbert space of this approach.
They are functions on the full gauge group $G$ defined on the graph dual to the triangulated boundary.
At every vertex the holonomies taken in a representation of $G$ ($\lambda_f=(j^+_f,j^-_f)$) are projected
to irreducible SU(2) representations ($j_{tf}$) and then coupled by SU(2) invariant intertwiners
($\iota^{\{ j_{tf_i}\} } (k_t)$). The dependence of the normals $x_t$ can be pushed from the matrix coefficients
into the intertwiners. Its only effect is that the projection is done on the ``rotated" subgroup
$H_{x_t}$ introduced after eq. \eqref{projxxx}.
We see that here the projected spin networks also appear as a basis of spin foam boundary states in
the perfect agreement with the canonical approach.

The coefficient $A(\lambda_f,j_{tf},k_t)$ in \eqref{reprampl} is nothing else but the vertex
amplitude of spin foam quantization in the spin network basis.
It contains the crucial information about dynamics of the theory.
It is in this place where the concrete form of the measure $\CD^{(x_0)} [\gl_t]$ becomes important.
Up to now the derivation worked well for any measure consistent with \eqref{trmeas}.
But without its knowledge we cannot evaluate the vertex amplitude.

There is actually another place where the explicit form of the measure is implicated.
We expect that the measure contains a $\delta$-function because the second class constraints
fix a half of components of the connection. From \eqref{simbs_init} one therefore
expects that some restrictions will arise on the group elements $g_f$. These restrictions are the same
discretized secondary second class constraints that have been imposed on $\gl_t$.
Their implementation corresponds in the canonical approach to the passage from the enlarged Hilbert
space consisting from {\it all} projected spin networks to the kinematical Hilbert space
with the second class constraints implemented at the quantum level \cite{AK,Alexandrov:2007pq}.
One expects that its effect is to fix some labels of the projected spin networks as,
for example, to restrict oneself to the sector with only {\it simple} representations.

\section{Closure constraint revisited}
\label{sec_conseq}

Usually the closure constraint \eqref{clcon} is considered as a quantization of a classical relation
between bi-vectors $B_f$ associated to triangles of a tetrahedron.
Since the triangles are not arbitrary, but form the tetrahedron, their bi-vectors must satisfy \cite{BCE}
\be
\vphantom{A\over B}
\smash{\sum\limits_{f\subset\, t}}B_{f}=0.
\label{classclos}
\ee
After quantization, $B_f$ are represented by generators $T_f$ acting on the representation spaces
assigned to the triangles. This gives the constraint \eqref{clcon} on the intertwiners which enter
boundary states and vertex amplitudes of spin foam models.

From our point of view this procedure is too naive to be able to capture possible corrections
which might appear at the quantum level. A more fair strategy would be to {\it derive} the closure constraint
from a path integral representation of a spin foam model.
Indeed, as was noticed in \cite{Freidel:2007py}, the closure constraint is imposed automatically
by integration over holonomy group elements in the partition function and thus it does not need to be
imposed by hand.\footnote{Although in \cite{Engle:2007qf} the closure constraint is imposed strongly
on the intertwiners of boundary states, it was also noticed that its classical counterpart \eqref{classclos}
can be obtained by the variation of the connection in the discrete action. This can be considered as another
indication in favor of treating the closure constraint as an equation of motion which can fluctuate at the quantum level.}
In other words, it is not an ingredient, but rather a consequence of the spin foam quantization.
This is consistent with the fact that its classical analogue, the Gauss constraint,
is first class and therefore it is generated simply by integration over the corresponding Lagrange
multiplier. In contrast, the second class constraints, supplied with an appropriate
determinant, must be inserted into the path integral measure from the very beginning
(see discussion in section \ref{sec_constraint}).

At the quantum level the closure constraint is nothing else but the invariance property of the intertwiners
entering the spin foam boundary states. If the basis in this state space is realized by the projected spin networks,
as in \eqref{reprampl}, the intertwiners can be read off from \eqref{projspin}.
Let $\Db_{pq}^{(\lambda_f)}(g_f)$ denote matrix coefficients of an element of the group $G$ in the representation $\lambda_f$.
Then the intertwiner coupling them at the vertex $t$
depends on the normal $x_t$ and can be presented as \cite{Alexandrov:2007pq}
\be
N^{(\{\lambda_{i}\},\{j_{i}\},k_t)}_{p_1 \cdots p_4}(x_t)=\mathop{\sum}\limits_{\ell_{j_1}\cdots\ell_{j_4}}
\iota^{\{ j_{i}\} }_{\{ \ell_{j_{i}} \} }(k_t) \prod_{i=1}^4 \Db^{(\lambda_i)}_{p_i \ell_{j_{i}}}(\gx_{x_t}).
\label{genBC}
\ee
where the indices $\ell_j$ label the basis of the subspace $H_{\rm SU(2)}^{(j)}$ appearing in the decomposition
of the representation $\lambda$ on the subgroup
\be
H_G^{(\lambda)}=\bigoplus\limits_{j} H_{\rm SU(2)}^{(j)}.
\label{decompH}
\ee
It is easy to see that the intertwiner \eqref{genBC} satisfies
\be
\mathop{\sum}\limits_{q_{1}\cdots q_{4}}
\(\prod_{i=1}^4 \Db^{(\lambda_i)}_{p_i q_i}(\gx)\) N^{(\{\lambda_{i}\},\{j_{i}\},k_t)}_{q_1 \cdots q_4}(x_t)
=N^{(\{\lambda_{i}\},\{j_{i}\},k_t)}_{p_1 \cdots p_4}(\gx\cdot x_t).
\label{invN}
\ee
The infinitesimal version of this transformation law gives precisely the relaxed closure constraint
\eqref{clconrel}.

It is clear that the usual invariance is spoiled due to the dependence of $x_t$.
An easy way to restore it is to integrate over these normals. This is what usually
done in spin foam models by inserting integrals $\int dx_t$ in the boundary states
\cite{Engle:2007qf,Engle:2007mu,Engle:2007wy}. Of course, this insertion produces $G$-invariant
intertwiners.

Thus, to decide which version of the closure constraint is the relevant one, we have to understand whether or not
one should integrate over the normals to tetrahedra.
For this it is necessary to consider gluing of several simplices.
In terms of the boundary states, the gluing is achieved by integrating over holonomies associated to common triangles.

Let us assume that we do insert the integration over $x_t$ in the simplex boundary states.
Then one immediately runs into two problems.
First, for a tetrahedron shared by two simplices, there will be two integrals over its normal.
This means that one actually introduced two normals to the same tetrahedron, one for each simplex.
This is inconsistent because the normals form a part of the $B$ field which associates unique data
to the elements of the simplicial decomposition.\footnote{The two normals cannot be interpreted as
the same normal seen from different reference frames related to the two simplices. We defined it
without any relation to the reference frame of the simplex which did not appear at all in our discretization.}

Second and may be more important, the integration over $x_t$ is inconsistent with the measure
on $g_f$ induced by the second class constraints. Indeed, as we discussed in the end of section \ref{subsec_sbs},
the presence of $\delta$-function in the measure $\CD^{(x_t)} [\gl_t]$ leads to some conditions
on $g_f$. As a result, the measure for these group elements will also be non-trivial and in particular
it should depend on the normals to the tetrahedra sharing the triangle $f$.
However, once the integral over $x_t$ has been inserted, one does not have them at our disposal,
the non-trivial measure cannot be written and the gluing becomes impossible.

Thus, we conclude that the integration over $x_t$ in the boundary states is inconsistent with the second class
constraints and the closure constraint must be imposed in its relaxed form \eqref{clconrel}.
Several remarks are in order:
\begin{itemize}
\item
The reasoning we gave above can be applied only in the presence of the second class constraints.
If they are absent, the measure $\CD^{(x_t)} [\gl_t]$ coincides with the usual one
and the dependence of the normals disappears from the very beginning.
Of course, the normals can be artificially introduced, but there is no way to measure them.
Thus, we are free to insert integrals over $x_t$ and find the standard closure constraint \eqref{clcon}.
This is precisely what happens in the models \cite{Livine:2007vk,Freidel:2007py,Livine:2007ya}
based on coherent states where the secondary constraints are ignored and the measure over
holonomies is taken as the standard Haar measure.
In this case it is independent of the normals so that the integration over $x_t$ and the gluing
are mutually consistent.

\item
There is another way to see why the dependence of the normals should be preserved in the presence of
the second class constraints and is auxiliary in the opposite case.
In canonical theory the vectors $x^I$ are conjugated to a part of the spin connection hidden in the Gauss constraint.
However, as we mentioned in section \ref{subsec_connect}, to quantize the theory one passes to another connection,
either $\SA_i^{IJ}$ or $\SSA_i^{IJ}$. Both modified connections commute with $x^I$ \cite{SAcon}.
The reason for this is that they have 3 independent components less comparing to
the spin connection. (They satisfy 9 second class constraints, whereas the spin connection fulfills only 6.)
The missing components are precisely the ones which are conjugated to $x^I$.
Therefore, these unit vectors should be added to the list of configuration variables.
In contrast, if there are no second class constraints, the connection used in holonomy operators
coincides with the usual spin connection and does not commute with $x^I$.

\item
One can notice that even when two simplices are glued together, to obtain the common boundary state,
it is not needed to integrate over the normal to the tetrahedron shared by the simplices.
Indeed, the gluing gives rise to a contribution to the boundary state which can be schematically written
as follows
\be
\Psi_{12}\[g_f;x_t\]
=\int \prod_{t_{12}}\CD^{(\xh_{t_{12}})} [\rhoh_{f_{12}}]\,
\Psi_1\[\gh_{f_1},\gh_{f_{12}}\rhoh_{f_{12}};\xh_{t_1},\xh_{t_{12}}\]
\Psi_2\[\gh_{f_2},\rho_{f_{12}}^{-1};\xh_{t_2},\xh_{t_{12}}\],
\label{glue_simplex}
\ee
where $f=(f_1,f_2,f_{12}),t=(t_1,t_2,t_{12})$, the labels 1,2 refer to non-shared faces and tetrahedra of
the corresponding glued simplices and the label 12 marks the shared faces and tetrahedron.
It is easy to see that, similarly to what happens in the vertex amplitude \eqref{amplit_simplex},
the dependence on $x_{t_{12}}$ drops out.
Moreover, an integral over $x_{t_{12}}$ would produce an infinite factor in the Lorentzian case.
Therefore, even in the complete partition function all normals should be kept fixed.
This corresponds to a gauge fixing of boosts in the discretized path integral.
Note that in the models proposed in \cite{Engle:2007qf,Freidel:2007py,Livine:2007ya} this issue does not arise
because all the ingredients, the boundary amplitudes to be glued and the measure to be used
for the gluing, are independent of the normals.

\item
The necessity to use the relaxed closure constraint resolves
a problem with the model proposed in \cite{Freidel:2007py}, which has been found in the work \cite{Engle:2007mu}.
It was noticed that in that model the space of intertwiners remains unrestricted despite the simplicity constraints
have been imposed. This is indeed true, but only when one considers the SO(4) invariant subspace
so that the correct statement is: the gauge-invariant projection of
the span of all intertwiners of \cite{Freidel:2007py} is equal to $\cI$, the space
of all SO(4) invariant intertwiners. In our language this originates from the fact that the intertwiners
\eqref{genBC} integrated with respect to $x_t$ form an over-complete basis in $\cI$ and hence choosing a subset of
these intertwiners does not necessarily reduce the projection of their span.
On the other hand, if one allows for dependence on the normals to tetrahedra,
the intertwiners satisfy the relaxed closure constraint and span a larger space $\cIr$.
In this case any restriction on the labels reduces to a certain subspace of $\cIr$ and one finds that
the model of \cite{Freidel:2007py} does put some constraints on these {\it covariant} intertwiners.

\item
Finally, we mention that the need for a modification of the relation between bi-vectors and generators of the gauge algebra,
which is closely related to the relaxation of the closure constraint \cite{Alexandrov:2007pq},
was observed in \cite{Oriti:2007vf}.
Besides, the intertwiners satisfying the relaxed closure constraint appeared also in
\cite{Livine:2005tp,Oriti_conf} in the group field theory context as well as in the case of coupling of particles.
Thus, they seem to be quite natural for the spin foam approach.
In this section we argued that in the presence of second class constraints they are actually needed
to describe the correct boundary state space.

\end{itemize}

\section{Example: SU(2) BF theory}

Since we do not know the correct discrete measure incorporating the secondary second class constraints
for gravity \eqref{contA}, our construction is somewhat inexplicit.
To highlight its various aspects, we illustrate it on a simple example, which is however closely related
to the model of \cite{Engle:2007qf} and LQG.

This example serves as a model for quantization based on the connection $\SSA_i^{IJ}$ relevant
to the Ashtekar--Barbero approach. In this case the secondary constraints are given by \eqref{su2con}.
They are much easier to discretize comparing to \eqref{contA} because they simply mean that,
for a constant $x^I$, $\SSA_i^{IJ}$
has vanishing boost part and its holonomies belong to an SU(2) subgroup.
Working in the Riemannian case, it is easy to write the corresponding measure as
\be
\CD^{(x_t)}[\gl_t ]=\delta\((\gl^+_{t})^{-1} \gl^-_{t}\, u_{x_t}\)\, d\gl_t^+ \, dg_t^-,
\label{measure_su2}
\ee
where $u_{x_t}=\gx_{x_t}^-(\gx_{x_t}^+)^{-1}\in {\rm SU(2)}$.

It is important to notice that in this case the new version of the simplicity constraints \eqref{newsim}
is inconsistent with the constraints on the connection because it leads to the vanishing
discretized action \eqref{dact}. There is however a similar, but inequivalent way to write the
initial quadratic constraints
\be
I_{(P)KL}^{IJ}(x)B^{KL}=0.
\label{newsim2}
\ee
It implies that only the ${\rm SU}_x(2)$ part of the $B$ field is non-trivial and corresponds
to the topological sector of Plebanski formulation where $B^{IJ}=e^I\wedge e^J$.
After discretization, the simplicity constraints can be written as
\be
\hat B_f^- = u_{x_{u(f)}}\,\hat B_f^+\, u_{x_{u(f)}}^{-1},
\label{opsim2}
\ee
where we have chosen the ``up" tetrahedron as the reference one (see the end of section \ref{subsec_discr}).

Inserting both second class constraints into the action \eqref{dact}, one finds that it reduces
to the discretized action of SU(2) BF theory. Thus, this example provides a quantization of this simple topological
theory embedded into an SO(4) covariant framework. The fact that we used constraints leading to Ashtekar--Barbero
formulation does not imply of course that its quantization is equivalent to the BF theory.
The reason is twofold. First, the connection $\SSA_i^{IJ}$, in contrast with $\SA_i^{IJ}$,
is not equal to the spin connection on the constraint surface. Hence, the
initial action expressed in terms of $\SSA_i^{IJ}$ will have a more complicated form than the action of BF theory.
And second, as we mentioned, $\SSA_i^{IJ}$ is not really a spacetime connection and therefore
it cannot be used in the spacetime covariant quantization: its path ordered exponentials cannot be interpreted as holonomies.

Despite the triviality of the example, it is instructive to see
how the construction described in the previous sections works.
First, the knowledge of the explicit form of the measure for
holonomies allows to evaluate the vertex amplitude \eqref{amplit_simplex}.
Substituting \eqref{measure_su2}, one obtains
\be
A(\lambda_f,j_{tf},k_t)=\int \prod_t dh_t\,\Psi^{(\lambda_f,j_{tf},k_t)}\[(\gx^{h_{d(f)}})^{-1}\gx^{h_{u(f)}};x_0\].
\label{amplit_SU2}
\ee
Using the property
\be
\mathop{\sum}\limits_{m,m',n,n'}\overline{\ClG{j^+}{j^-}{j_1^{\vphantom{+}}}{m^{\vphantom{+}}}{m'}{\ell_1}}
D^{(j^+)}_{mn^{\vphantom{+}}}(h)D^{(j^-)}_{m'n'}(h) \ClG{j^+}{j^-}{j_2}{n^{\vphantom{+}}}{n'}{\ell_2}
=\delta_{j_1j_2} d_{j_1} D^{(j_1)}_{\ell_1 \ell_2}(h),
\label{property}
\ee
it is easy to realize that the vertex amplitude is given by
\be
A(\lambda_f,j_{tf},k_t)=\(\prod_f\delta_{j_{u(f)f}^{\vphantom{+}}j_{d(f)f}^{\vphantom{+}}}\)15J(j_{tf};k_t).
\label{amplit_SU2res}
\ee
Thus, the result indeed coincides with the vertex amplitude of SU(2) BF theory.

Notice that the dependence of the vertex amplitude on
the SO(4) representations $\lambda_f$ completely disappears and the SU(2) representations
$j_{u(f)f}$ and $j_{d(f)f}$ are required to coincide.
The same of course should be true for the boundary states.
The conditions on the labels of the projected spin networks
can be obtained taking into account the constraints on $g_f$ following from
\eqref{simbs_init} and \eqref{measure_su2}.
It follows that these group elements must satisfy
\be
g_f^-=u_{x_{u(f)}}\, g^+_{f}\, u_{x_{d(f)}}^{-1}.
\label{congf}
\ee
As a result, the element $G_f$ appearing in \eqref{projspin} belongs to the diagonal SU(2) subgroup
and the same property \eqref{property} can be used to get the restrictions.
The resulting boundary state coincides with the usual SU(2) spin network as it should be for SU(2)
BF theory.

This procedure gives an example of the explicit realization of imposing second class constraints
at the level of the Hilbert space. Moreover, precisely this example was employed in \cite{AlLiv}
to demonstrate that the Lorentz covariant
loop quantization based on $\SSA_i^{IJ}$ reproduces the kinematical Hilbert space of LQG.
This is not surprising because the kinematical Hilbert spaces of LQG and SU(2) BF theory are isomorphic.

Notice that if one integrates the simplex boundary state over $x_t$, the reduction to the SU(2) spin networks
would become impossible.
The reason is that the normals appearing in the boundary state and in the constraint \eqref{congf} would be
decoupled: the first one is integrated over and the second is fixed. As a result, they would not cancel and
the argument of the projected spin networks would not reduce to
the diagonal subgroup.\footnote{In \cite{Engle:2007qf} the restrictions on representations were
achieved by weakly imposing the simplicity constraints
on the $B$ field. In our approach the constraints on $B_f$
are already taken into account by the explicit $\delta$-function in the measure. Besides, although
the method of \cite{Engle:2007qf} leads to spin networks with the same set of labels as that of LQG, the states
of the two models are {\it physically} different
because the former depend on SO(4) holonomies and the latter are SU(2) spin networks.}
This shows the importance of keeping the normals $x_t$ as free arguments of the boundary states
and, as a consequence, of the use of the relaxed closure constraint.

\section*{Acknowledgements}

The author is grateful to Kirill Krasnov and Philippe Roche for interesting discussions.
This research is supported by CNRS and by the contract ANR-06-BLAN-0050.


\begin{thebibliography}{99}


\bibitem{Oriti}
D.~Oriti, ``Spacetime geometry from algebra:
spin foam models for non-perturbative quantum gravity,"
Rept.\ Prog.\ Phys.\ {\bf 64}, 1489 (2001)
[gr-qc/0106091].

\bibitem{perez}
A.~Perez, ``Spin foam models for quantum gravity,"
Class.\ Quant.\ Grav.\  {\bf 20}, R43 (2003)
[gr-qc/0301113].


\bibitem{PonzanoRegge}
G. Ponzano and T. Regge.
``Semiclassical limit of Racah coefficients." in {\it Spectroscopy and Group Theoretical
Methods in Physics} ed. by F. Bloch (North-Holland, Amsterdam, 1968).

\bibitem{Noui:2004iy}
K.~Noui and A.~Perez,
``Three dimensional loop quantum gravity: Physical scalar product and  spin
foam models,"
Class.\ Quant.\ Grav.\  {\bf 22} (2005) 1739
[arXiv:gr-qc/0402110].

\bibitem{Plebanski:1977zz}
J.~F.~Plebanski,
``On the separation of Einsteinian substructures,"
J.\ Math.\ Phys.\  {\bf 18} (1977) 2511.

\bibitem{De Pietri:1998mb}
R.~De Pietri and L.~Freidel,
``so(4) Plebanski Action and Relativistic Spin Foam Model,"
Class.\ Quant.\ Grav.\  {\bf 16} (1999) 2187
[arXiv:gr-qc/9804071].

\bibitem{Freidel:1998pt}
L.~Freidel and K.~Krasnov,
``Spin foam models and the classical action principle,"
Adv.\ Theor.\ Math.\ Phys.\  {\bf 2}, 1183 (1999)
[hep-th/9807092].


\bibitem{BCE}
J.W.~Barrett and L.~Crane,
``Relativistic spin networks and quantum gravity,"
J. Math. Phys. {\bf 39}, 3296 (1998)
[gr-qc/9709028].

\bibitem{BC}
J.W.~Barrett and L.~Crane,
``A Lorentzian signature model for quantum general relativity,"
Class. Quantum Grav. {\bf 17}, 3101 (2000)
[gr-qc/9904025].



\bibitem{Livine:2007vk}
E.~R.~Livine and S.~Speziale,
``A new spinfoam vertex for quantum gravity,"
Phys.\ Rev.\  D {\bf 76} (2007) 084028
[arXiv:0705.0674 [gr-qc]].


\bibitem{Engle:2007uq}
J.~Engle, R.~Pereira and C.~Rovelli,
``The loop-quantum-gravity vertex-amplitude,"
Phys.\ Rev.\ Lett.\  {\bf 99} (2007) 161301
[arXiv:0705.2388 [gr-qc]].


\bibitem{Alexandrov:2007pq}
S.~Alexandrov,
``Spin foam model from canonical quantization,"
Phys.\ Rev.\  D {\bf 77} (2008) 024009
[arXiv:0705.3892 [gr-qc]].


\bibitem{Engle:2007qf}
J.~Engle, R.~Pereira and C.~Rovelli,
``Flipped spinfoam vertex and loop gravity,"
arXiv:0708.1236 [gr-qc].

\bibitem{Freidel:2007py}
L.~Freidel and K.~Krasnov,
``A New Spin Foam Model for 4d Gravity,"
arXiv:0708.1595 [gr-qc].

\bibitem{Livine:2007ya}
E.~R.~Livine and S.~Speziale,
``Consistently Solving the Simplicity Constraints for Spinfoam Quantum Gravity,"
arXiv:0708.1915 [gr-qc].

\bibitem{Oriti:2007vf}
D.~Oriti and T.~Tlas,
``A New Class of Group Field Theories for 1st Order Discrete Quantum Gravity,"
Class.\ Quant.\ Grav.\  {\bf 25} (2008) 085011
[arXiv:0710.2679 [gr-qc]].

\bibitem{Engle:2007mu}
J.~Engle and R.~Pereira,
``Coherent states, constraint classes, and area operators in the new
spin-foam models,"
arXiv:0710.5017 [gr-qc].

\bibitem{Pereira:2007nh}
R.~Pereira,
``Lorentzian LQG vertex amplitude,"
arXiv:0710.5043 [gr-qc].

\bibitem{Engle:2007wy}
J.~Engle, E.~Livine, R.~Pereira and C.~Rovelli,
``LQG vertex with finite Immirzi parameter,"
arXiv:0711.0146 [gr-qc].



\bibitem{Buffenoir:2004vx}
E.~Buffenoir, M.~Henneaux, K.~Noui and Ph.~Roche,
``Hamiltonian analysis of Plebanski theory,"
Class.\ Quant.\ Grav.\  {\bf 21} (2004) 5203
[gr-qc/0404041].

\bibitem{ABR}
S.~Alexandrov, E.~Buffenoir, and Ph.~Roche,
``Plebanski Theory and Covariant Canonical Formulation,"
[gr-qc/0612071].


\bibitem{Henneaux:1994jf}
M.~Henneaux and A.~Slavnov,
``A Note on the path integral for systems with primary and secondary second
class constraints,"
Phys.\ Lett.\  B {\bf 338} (1994) 47
[arXiv:hep-th/9406161].


\bibitem{SA}
S.~Alexandrov,
``SO(4,C)-covariant Ashtekar--Barbero gravity and
the Immirzi parameter,"
Class.\ Quantum\ Grav.\ {\bf 17}, 4255 (2000)
[gr-qc/0005085].

\bibitem{AV}
S.~Alexandrov and D.~Vassilevich,
``Area spectrum in Lorentz covariant loop gravity,"
Phys.\ Rev.\ D {\bf 64}, 044023 (2001) [gr-qc/0103105].

\bibitem{SAcon}
S.~Alexandrov,
``On choice of connection in loop quantum gravity,"
Phys.\ Rev.\ D {\bf 65}, 024011 (2002)
[gr-qc/0107071].

\bibitem{AlLiv}
S.~Alexandrov and E.R. Livine,
``SU(2) loop quantum gravity seen from covariant theory),"
Phys. Rev. D {\bf 67}, 044009 (2003)
[gr-qc/0209105].

\bibitem{Livine:2006ix}
E.~R.~Livine,
``Towards a covariant loop quantum gravity,"
[gr-qc/0608135].



\bibitem{Batalin:1981jr}
I.~A.~Batalin and G.~A.~Vilkovisky,
``Gauge Algebra And Quantization,"
Phys.\ Lett.\  B {\bf 102} (1981) 27.

\bibitem{Batalin:1984jr}
I.~A.~Batalin and G.~A.~Vilkovisky,
``Quantization Of Gauge Theories With Linearly Dependent Generators,"
Phys.\ Rev.\  D {\bf 28} (1983) 2567
[Erratum-ibid.\  D {\bf 30} (1984) 508].


\bibitem{Batalin:1997dy}
I.~A.~Batalin, K.~Bering and P.~H.~Damgaard,
``Second class constraints in a higher-order Lagrangian formalism,"
Phys.\ Lett.\  B {\bf 408} (1997) 235
[arXiv:hep-th/9703199].


\bibitem{Ashtekar:1986yd}
A.~Ashtekar,
``New Variables For Classical And Quantum Gravity,"
Phys.\ Rev.\ Lett.\  {\bf 57} (1986) 2244.

\bibitem{Ashtekar:1991hf}
A.~Ashtekar,
``Lectures on nonperturbative canonical gravity," {\it notes prepared in collaboration with R. Tate}
(World Scientific, Singapore, 1991).

\bibitem{Immirzi:1992ar}
G.~Immirzi,
``The Reality conditions for the new canonical variables of general relativity,"
Class.\ Quant.\ Grav.\  {\bf 10} (1993) 2347
[arXiv:hep-th/9202071].

\bibitem{Alexandrov:2005ng}
S.~Alexandrov,
``Reality conditions for Ashtekar gravity from Lorentz-covariant formulation,"
Class.\ Quant.\ Grav.\  {\bf 23} (2006) 1837
[arXiv:gr-qc/0510050].


\bibitem{Barbero:1994an}
J.~F.~Barbero,
``Reality conditions and Ashtekar variables: A Different perspective,"
Phys.\ Rev.\ D {\bf 51} (1995) 5498
[gr-qc/9410013].


\bibitem{psn}
E.R. Livine,
``Projected Spin Networks for Lorentz connection:
Linking Spin Foams and Loop Gravity,"
Class. Quantum Grav. {\bf 19}, 5525 (2002)
[gr-qc/0207084].


\bibitem{AK}
S.~Alexandrov and Z.~Kadar,
``Timelike surfaces in Lorentz covariant loop gravity and spin foam models,"
Class.\ Quant.\ Grav.\ {\bf 22}, 3491 (2005)
[gr-qc/0501093].


\bibitem{Livine:2005tp}
E.~R.~Livine and D.~Oriti,
``Coupling of spacetime atoms and spin foam renormalisation from group field theory,"
JHEP {\bf 0702} (2007) 092
[arXiv:gr-qc/0512002].


\bibitem{Oriti_conf}
D.~Oriti,
``Group field theory and simplicial quantum gravity," talk at
QG2 2008 Quantum Geometry and Quantum Gravity Conference.


\end{thebibliography}
\end{document}